\documentclass[]{article}

\usepackage{times}
\usepackage{amsmath,amssymb}
\usepackage{bm}

\usepackage{graphicx}
\graphicspath{{./figs/}}
\usepackage{url}
\usepackage{color}

\usepackage[section]{placeins}
\usepackage{booktabs}
\usepackage{rotating}

\title{Urban spatial structures from human flow by Hodge-Kodaira decomposition}

\author
{Takaaki Aoki,$^{1\ast}$ Shota Fujishima,$^{2}$ Naoya Fujiwara$^{3,4,5}$\\
\\
\normalsize{$^{1}$Faculty of Education, Kagawa university, Takamatsu 760-8521, Japan}\\
\normalsize{$^{2}$Graduate School of Economics, Hitotsubashi University, Tokyo 186-8601, Japan}\\
\normalsize{$^{3}$Graduate School of Information Sciences, Tohoku University, Sendai 980-8579, Japan}\\
\normalsize{$^{4}$Center for Spatial Information Science, the University of Tokyo, Kashiwa 277-8568, Japan}\\
\normalsize{$^{5}$Institute of Industrial Science, the University of Tokyo, Tokyo 153-8505, Japan}
\\
\normalsize{$^\ast$E-mail:  takaaki.aoki.work@gmail.com.}
}

\date{} %

\begin{document}

\maketitle 

\begin{abstract}
Human flow in cities indicates social activity and can reveal urban spatial structures based on human behaviours for relevant applications.
Scalar potential is a mathematical concept, and if successfully introduced, it can provide an intuitive perspective of  human flow.
However, the definition of such a potential to the origin-destination flow matrix and determination of its plausibility remain unsolved.
Here, we apply Hodge-Kodaira decomposition, in which a matrix is uniquely decomposed into a potential-driven (gradient) flow and a curl flow.
We depict the potential landscapes in cities due to commuting flow and reveal how the landscapes have been changed or unchanged by years or transport methods.
We then determine how well the commuting flow is described by the potential, by evaluating the percentage of the gradient component for metropolitan areas in the USA and show that the gradient component is almost 100\% in several areas; in other areas, however, the curl component is dominant, indicating the importance of circular flow along triangles of places.
The potential landscape  provides an easy-to-use visualisation tool to show the attractive places of human flow and will aid in various applications in  commerce, urban design, and epidemic spreading.
\end{abstract}

\section*{Introduction}
\begin{figure}[tbp]
  \centering
  \includegraphics[width=\linewidth]{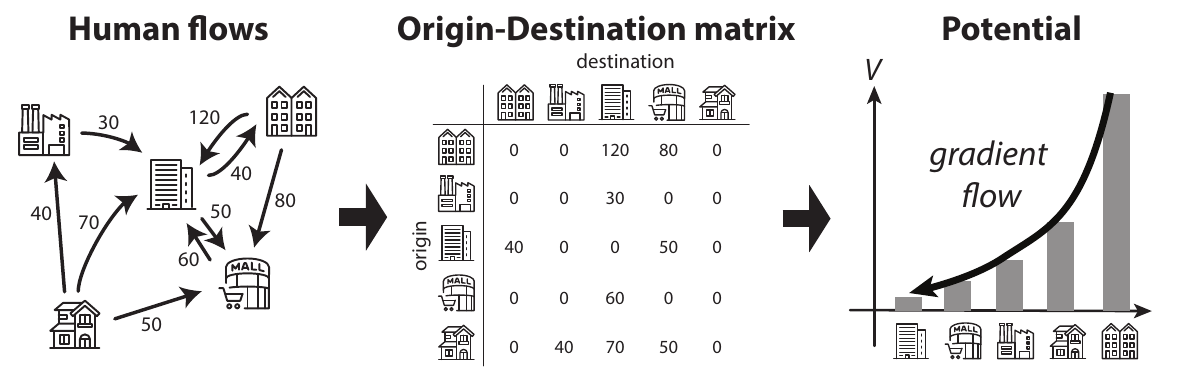}
  \caption{\textbf{Origin-Destination (OD) matrix and the concept of the potential of human flow.}}
  \label{fig:odtable}
\end{figure}

Human mobility is a vital social activity in our society that is relevant to various applications in commerce, urban design, marketing, and economics while also being involved in the spreading of diseases such as COVID-19.
Today, the popularisation of mobile phones has enable the collection of real-time, massive data, in addition to ordinary survey-based collection.
The collected data are usually aggregated as an \textit{origin-destination (OD) matrix} (Fig. \ref{fig:odtable}),
which describes how many people are moving from one place (origin) to another (destination).
Thus, the mobility data characterises the relationships between places based on human behaviour,
and is expected to reveal the places that attract human flow and their basins.
Such information tells us the centres and limits of cities based on the human behaviour of real-time movements and unfolds the actual shapes of cities,
which is dynamically changeable according to years, transportation methods, and movement restrictions.
They, in turn, aid location decision making for commercial or public buildings, the optimisation of transportation systems, urban planning by policy makers, and the measures for movement restrictions in COVID-19 spreading.

To reveal the spatial structure of cities, 
we consider the {\it scalar potential} of human flow.
Potential is a popular mathematical concept used in various scientific fields, from physics to economics.
In the context of our study, it is defined as a function of location, and its gradient yields a net movement of people among locations.
Such a potential landscape provides an intuitive perspective of human flow by analogously representing water flowing from a higher place to a lower place.
Furthermore, it reduces the relational flow data to location-level statistics that are ready to be shown on a map.
From the map, we can easily identify the sinks and sources of human flow, as illustrated in Fig. \ref{fig:odtable}. 
A sink of human flow indicates attractive places. 
Potential landscape can visualise the urban structure behind massive data of human mobility
and utilise it for relevant applications, if successfully introduced.

However, it is not obvious how to introduce the potential to human flow.
Unlike an electromagnetic field,
human flow is not described by a two-dimensional vector field, but as an OD matrix.
Furthermore, it is an open question whether human flow can be effectively described by a potential in the first place, according to Helmholtz's theorem.
In the literature, 
the OD matrix is converted to a 2D vector field by averaging all trips from each location \cite{Mazzoli2019b}.
This is analogous to considering only the motion of the centre of mass instead of the motions of the individuals.
The resultant vector field was found to be almost irrational, and a scalar potential was introduced.
However, this aggregation discards the place-to-place information of the original data.
As demonstrated in benchmark tests using synthetic data (Supplementary Note 1),
it is difficult using the previous method to identify the number of centres and their areas expressed in the given data.

Another approach is to define a potential \cite{Stewart1947} or an attractiveness \cite{HarrisWilson1978} based on the so-called gravity model \cite{gravity_zipf1946,Ullman1956,WilsonBook1974}, which is a well-known model of human flow.
These measures have been evaluated by using several residential and economic datasets \cite{Geurs2004,Ellam2018}.
However, these measures are specific to the assumed model and are not calculated from the OD matrix data.
Here, we provide a straightforward introduction of a potential to the OD matrix by applying the Hodge-Kodaira decomposition of graph flow \cite{DeRham1984,hodge1989theory,Jiang2011,Kodaira1949,Warner1983}.
As described in the Method section, the human flow is uniquely decomposed into a potential-driven (gradient) flow and another circular flow.
The potential at each place is directly and easily calculated from a given OD matrix
without any model assumptions and calibration parameters.
The potential is interpretable: it refers to the difference between night-time and daytime populations for commuting trips.
Furthermore, the decomposition allows us to determine how well the potential describes human flow by evaluating the percentage of the gradient component.
We find that the circular component in human flow is not always negligible.
This is in contrast to  previous studies where flow is, by assumption, described as the gradient component \cite{Stewart1947} or the circular flow is treated as noise \cite{Mazzoli2019b}.

In the following, we depict the potentials of the commuting flow in London for several different transport methods,
and show the evolution of the potential landscape over 30 years in Tokyo. 
We then study the percentage of the gradient component in metropolitan areas in the USA. 
Finally, we discuss the practical meaning of the potential and limitations of the method.

\section*{Overview of Hodge-Kodaira decomposition to an OD matrix}
Here, we overview  Hodge-Kodaira decomposition, which is applied to an OD matrix (see Methods for the details).
First, we consider the net flow of movement from a given  OD matrix $M$ as
when 150 persons move from location $i$ to another location $j$ and 50 people move in the opposite direction, we consider the net movements of 100 persons from $i$ to $j$.
The net flow is given by
\begin{equation}
  A = M - M^{\intercal}, \label{eq:netflow}
\end{equation}
where $M^{\intercal}$ denotes the transpose of $M$.
Matrix $A$ is skew-symmetric, that is, $A_{ij} = - A_{ji}$,
and is possibly described by combinatorial gradient of a potential $s$, given by
\begin{equation}
  (\text{grad}\, s)(i, j) =  s_j-s_i.
\end{equation}

Then, we define an optimisation problem for potential $s$:
\begin{equation}
  \min_s  \lVert   \text{grad}\ s - A \rVert_2 = \min_s \left[ \sum_{i,j} \left[ (s_j - s_i) - A_{ij} \right]^2 \right]. \label{eq:optimization_problem}
\end{equation}
According to the combinatorial Hodge theory \cite{Jiang2011}, the space of net flow $\mathcal{A}$ is orthogonally decomposed into two subspaces:
\begin{equation} 
  \mathcal{A}  = \text{im}(\text{grad})  \oplus \text{im}(\text{curl}^*),
\end{equation}
where $\text{curl}$ is the combinatorial curl operator and $\text{curl}^*$ is its adjoint operator.
Thus, the optimisation problem is equivalent to an $l_2$-projection of $A$ onto im(grad),
and the minimal norm solution is simply given by
\begin{equation}
  s_i = -\frac{1}{N} \text{div} A = - \frac{1}{N} \sum_{j=1}^N A_{ij}, \label{eq:potential}
\end{equation}
where $s_i$ is the potential at the $i$th location and $N$ is the number of locations. 
It is noted that $s_i$ is negative potential ($s_i=-V_i$).
This means that we observe more trips from a place with lower potential to another place with higher potential.

\section*{Potential landscapes in cities}
\begin{figure}[tb]
  \centering
  \includegraphics[width=12cm]{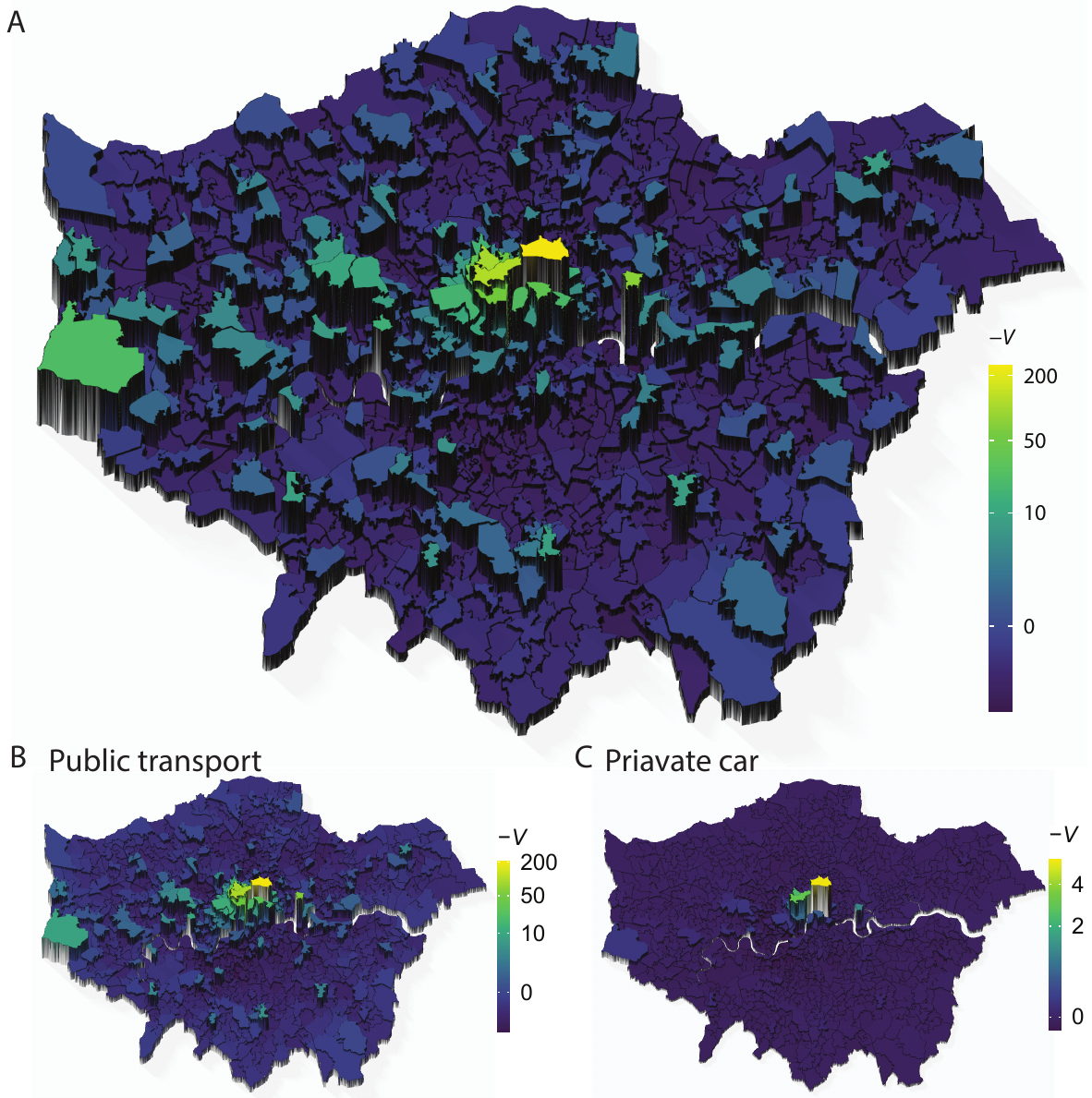}
  \caption{
    \textbf{Negative potential $-V$ of the home-work trips in London.}
    (\textbf{A}) 
    The potential at a place is indicated by its colour and its height.
    (\textbf{B,C})
    The potentials are depicted for the selected trips by specific transport methods. 
  }
  \label{fig:PotentialLondon}
\end{figure}

Figure \ref{fig:PotentialLondon} depicts the negative potential $-V_i (=s_i)$ of the decomposed gradient flow in Greater London, using a person-trip dataset in 2011.
The OD matrix shows the number of commuters aggregated by the middle layer super output area (MSOA) in the census 2011.
The potential has the largest peak at ``City of London 001'', literally the centre of London.
Its neighbouring areas, such as ``Westminister 018'' and ``Westminister 013'', also have a large potential.
Another peak, that is, a local maximum in the potential landscape,  is seen at ``Tower Hamlets 033'',
and there are small peaks outside the central area of London.
Most other  areas are characterised by a relatively lower potential by $-V$, serving as the sources of commuters to the centres.

The flows selected by specific transport methods provide another picture of potential landscapes in London.
The potential for public transportation (Fig. \ref{fig:PotentialLondon}B) is rather similar to that by all methods.
In contrast, the potential for private cars (Fig. \ref{fig:PotentialLondon}C) becomes a \textit{single-center} city than \textit{multi-center}:
a few locations still have higher potential, and the other locations have very low potentials without small peaks.
In addition, the potential amplitude is smaller than other cases, reflecting the volume of commuters (public transport = 1.6 million trips , private car = 45.2 thousand trips).

\begin{figure}[tbp]
  \includegraphics[width=\linewidth]{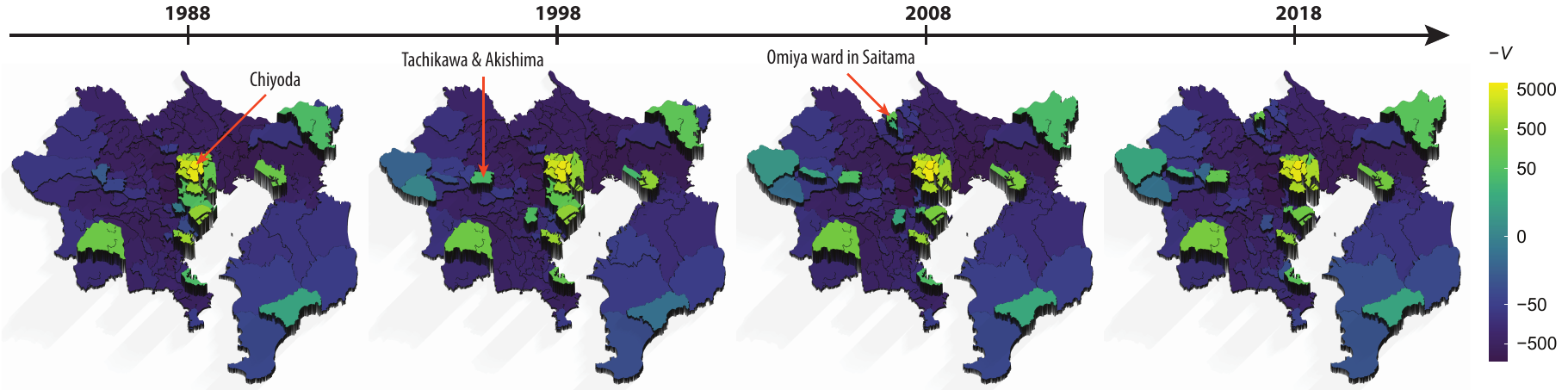}
  \caption{
    \textbf{Time-evolution of potential landscape  in Tokyo metropolitan area.} 
    The potentials are obtained from the home-work trips of successive surveys over 30 years in the Tokyo metropolis and surrounding provinces.
    Each  zone is basically equal to a municipal district.
  }
  \label{fig:PotentialTokyo}
\end{figure}

Next, we examined the time evolution of the potential landscape (Fig. \ref{fig:PotentialTokyo}),
using the commuter datasets of successive person-trip surveys from 1988 to 2018 in the Tokyo metropolitan area.
Over 30 years, \textit{Chiyoda} city --- the Imperial Palace and its surrounding areas ---  has been at the top of the potential.
The city is known as the economic and political centre in Japan:
it houses the headquarters of major enterprises, government institutions, and the Tokyo Central Railway Station.
Its neighbouring cities, such as \textit{Minato}, \textit{Chuo}, \textit{Shinjuku}, and \textit{Shibuya}, have occupied the top five ranks by potential over the years (Table \ref{tbl:top20_tokyo}), and formed the largest stable peak in the Tokyo metropolitan area.
Several small, steady peaks were observed outside the central area (e.g. \textit{Yokohama},  \textit{Chiba},  \textit{Kawasaki}, and \textit{Atsugi} cities).
In contrast to these steady peaks, new peaks appeared in \textit{Tachikawa} and \textit{Akishima} cities after 1998 and at the \textit{Omiya} ward in \textit{Saitama} city after 2008.
These small peaks correspond to the business cores envisioned by the fourth National Capital Regional Development Plan in 1986 \cite{Itsuki2006}, which aimed at multi-nucleated urban structures to avoid over-concentration in the Tokyo central area.
The potential over 30 years reveals how the urban structure in Tokyo has  changed or remained unchanged.

\section*{How much percentages of human flows is represented by the potential?}
\begin{figure}[tbp]
  \includegraphics[width=\linewidth]{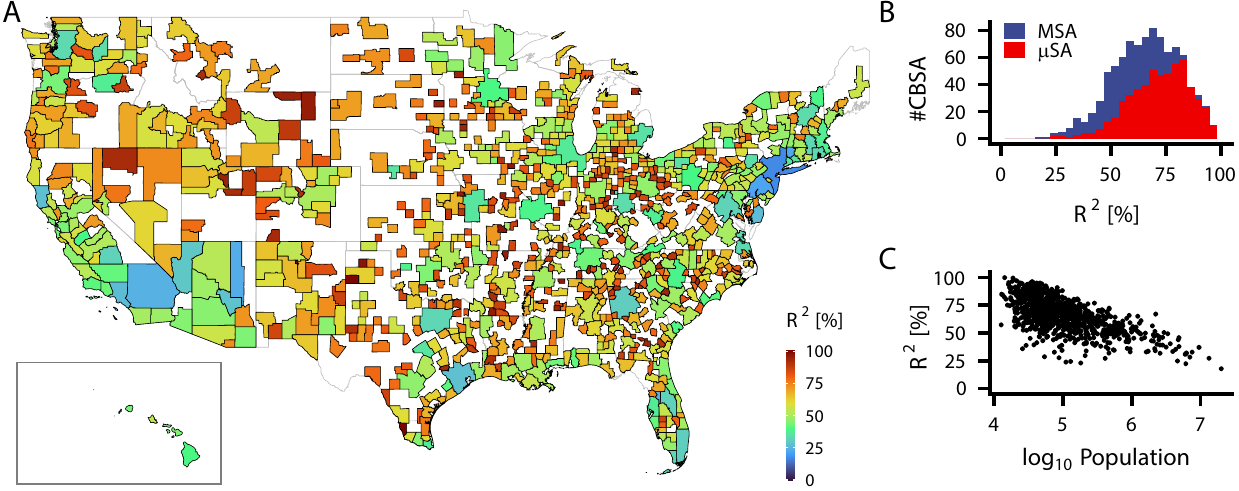}
  \caption{
  \textbf{The percentage of gradient component $R^2$ in human flow}.
    (\textbf{A}) Percentage $R^2$ in home-work trips for each Core-based statistical area (CBSA) in 2018. The inset shows the CBSAs in Hawaii.
    (\textbf{B}) Histogram of percentage $R^2$. CBSAs are classified into metropolitan statistical areas (MSAs) and micropolitan statistical areas ($\mu$SAs).
    (\textbf{C}) Percentage $R^2$ is plotted against the population of the CBSAs.
  }
  \label{fig:Percentage}
\end{figure}

To answer this question, we first define the percentage of the gradient component as
\begin{equation}
  R^2  =\frac{\sum_{i,j} \left[ (\text{grad}\, s)(i, j) \right]^2}{\sum_{i,j} A_{ij}^2} = 1 -  \frac{ \sum_{i} \sum_{j \neq i}  \left( A_{ij} - (s_j - s_i) \right)^2 }{\sum_{i} \sum_{j \neq i}  \left( A_{ij} \right)^2 }.
\end{equation}
This quantity is known as the so-called `coefficient of determination' in statistics, and 
is a reasonable choice for evaluating the explanatory power of the potential,
which is determined via orthogonal projection as ordinary least squares.

We evaluated the percentage $R^2$ for the metropolitan areas in the USA (Fig. \ref{fig:Percentage}A),
using the home-work trip datasets for each Core-based statistical area (CBSA) in 2018.
The percentage $R^2$ widely varies among the areas:
the minimal percentage $R^2$ = 17.72\% was in \textit{New York-Newark-Jersey City, NY-NJ-PA}, while the maximum was 99.98\% in \textit{Zapata, TX}. 
The distribution  has a mean of $\mu$ = 66.2\% and  standard deviation $\sigma$ = 15.3\% (Fig. \ref{fig:Percentage}B).
In CBSAs, metropolitan statistical areas (MSAs) tend to have a lower percentage than micropolitan statistical areas ($\mu$SAs).
Thus, the percentage $R^2$ is plotted against the population (Fig. \ref{fig:Percentage}C), showing that the percentage $R^2$ tends to decline for larger populations.
In addition, the percentage $R^2$ was changed by the transport methods in the London case (Table \ref{tbl:LondonPercentage}) and  by years in the Tokyo case (Table \ref{tbl:TokyoPercentage}).

\section*{Discussion}
\label{sec:discussion}
In this paper, we have introduced a potential for the OD matrix by  Hodge-Kodaira decomposition,
and depicted the potential landscape in cities.
In London, the largest peak of the potential landscape, that is, the most attractive centre of the flow, is located at  ``City of London 001''.
The landscape could give a different view of urban structure by the transportation method.
In the Tokyo metropolitan area, the time evolution of the potential over 30 years revealed how Tokyo has changed or remained unchanged in the view of human flow.
We found that the largest peak was stably located in \textit{Chiyoda} city, which is the central area of Tokyo.
Other peaks were seen at suburban business cores,
confirming the development of the multi-nucleated urban structure as envisioned in the national development plan in 1986.
These business cores are also known as ``edge cities'' of Tokyo, which are dynamically organised \cite{Garreau1991, Li2018, fujita1982multiple}.
In fact, it is clearly shown that some cores have emerged during the years as new peaks in the potential landscape.

We first discuss the practical meaning of the potential and how it is related to the spatial concentration of employment or daytime population, which have been used to depict polycentric intra-urban structures \cite{van2016pacifying, barthelemy2016structure}.
Using equations (\ref{eq:netflow}) and (\ref{eq:potential}), the potential is clearly interpreted as the balance between incoming and outgoing flows of people, mathematically given by,
\begin{equation}
  s_i =  \frac{1}{N} \left( \sum_j^N M_{ji} - \sum_j^N M_{ij} \right).
\end{equation}
The terms represent the incoming and outgoing flow, respectively.
Thus, the potential is the difference between the daytime and  night-time populations when the OD matrix describes the home-work trips of all residents.
The daytime population, that is, the population at destination places in home-work trips, is the first part of the potential, but not the entirety. The night-time population as the population at the origin is subtracted. 

The potential of human flow also incorporates the importance of circular flow in cities.
We evaluated how well the potential describes human flow in metropolitan areas in the US by using the percentage of the gradient component.
We found that the percentage is not always 100\% and not a universal value
but highly variable among the areas.
For several areas, the gradient component is more dominant, with a high percentage $R^2$,
while in other areas,  the other component (curl component) is dominant.
This variation reflects the differences in human flow across the areas
and raises new questions of when and why human flow is well described by the potential. %
Furthermore, the curl component tends to be dominant for large cities,
indicating the importance of circular flow in net movements of people.
The curl is defined for triplets of locations as described in Methods.
In contrast, human flow has been discussed in terms of paired locations: origin and  destination.
The circulation along triangle places addresses a new aspect of human flow with another question: What drives the circular flows in populated areas?
The decomposition method opens up new research avenue on human mobility and urban structures.

The limitation of the proposed method should be noted.
The potential is based on the rigorous mathematical definition of the OD matrix and 
does not require any model assumptions and any additional datasets.
Conversely, the analysis in this study does not consider several factors assumed in spatial interaction models, such as the gravity model \cite{gravity_zipf1946,Ullman1956,WilsonBook1974} or radiation model \cite{Simini2012}.
In particular, the distance deterrence on human mobility is not considered.
This could impose limitations on a native application  to a dataset at the country level, where distance  critically matters.
Thus, it is appropriate to apply decomposition to the human flow dataset within cities or narrow regions.
Otherwise, a distance-weighted function  can be integrated into the decomposition, as described in Supplementary Note 4.

In summary, 
the potential landscape by Hodge-Kodaira decomposition provides an intuitive perspective of human flow by its gradient flow from a higher place to a lower place.
The landscape allows us  to understand the spatial structure of cities based on  human movements rather than administrative circumstances
and to study the dynamic changes in the spatial structure under different conditions.
For example, we can study whether the global increase in remote workers due to the COVID-19 pandemic is alleviating over-concentration of population in city centres by checking the emergence of new potential peaks in suburbs or the decline of preexisting ones. 
The method  provides an easy-to-use visualisation tool to show the attractive places of  human flow and will aid relevant applications  in commerce, urban design, and epidemic spreading.

\bibliographystyle{plain}
\bibliography{../ref,../ref_datasets}

\section*{Acknowledgments}
We thank S. Segi,  T. Mori, R. Lambiotte and S. Shinomoto for fruitful discussions.

\section*{Funding}
This work was supported by the Research Institute for Mathematical Sciences, a joint research centre at Kyoto University(TA);
JSPS KAKENHI Grant Number JP18K12776 (SF);
JSPS KAKENHI Grant Number JP21H03507 (NF).

\section*{Author contribution}
All authors designed the study, discussed the implications of the data analysis, and wrote the manuscript.
TA implemented the method. 
TA and SF performed the data analysis of London data (TA), CBSAs in the US (SF,TA), and Tokyo (TA).

\section*{Competing interests}
There are no potential conflicts of interest among the authors.

\section*{Data availability}
The person trip datasets and the census data that support the findings of this study are publicly available, as noted in the Methods.

\section*{Code availability}
The code is available in the GitHub repository at \url{https://github.com/TakaakiAokiWork/HodgePotentialHumanFlow/}.

\section*{Supplementary Materials}
\section*{Methods}
\subsection*{Hodge-Kodaira decomposition to OD matrix}

The origin-destination (OD) matrix $M$ is a square matrix that represents the number of trips from origin $i$ to destination $j$ by its elements $M_{ij}$.
Any square matrix $M$ is uniquely decomposed into a symmetric and a skew-symmetric matrix,
\begin{align}
  M = \underbrace{\frac{1}{2} \left( M + M^{\intercal} \right) }_{\text{symmetric }}
  + \underbrace{\frac{1}{2} \left( M - M^{\intercal} \right) }_{\text{skew-symmetric }},
  \label{eq:symm}
\end{align}
where $M^{\intercal}$ is the transpose of $M$.
The symmetric part can be further decomposed into diagonal and off-diagonal elements.
The former represents a self-loop flow at each location.
The latter part is the bidirectional circulation of people among two locations.
Although it would be interesting to look into these symmetric elements,
we focus here on the skew-symmetric part because it is possibly described by the gradient flow.
In this study, we analyse  $A$ ($= M - M^{\intercal}$) by multiplying the skew-symmetric part by 2.
Matrix $A$ represents the net movements of human flow, removing the self-loop and bidirectional circulations.

We decompose the net flow $A$ by the Hodge-Kodaira decomposition.
The decomposition is, in general, defined to an undirected graph $G(\mathcal{V},E)$,
where the vertex set is $\mathcal{V}$ and the edge set is $E$.
The element  $A_{ij}$ represents the flow at the edge \cite{Jiang2011}.
According to \cite{Jiang2011}, the combinatorial gradient operator and  combinatorial curl operator are defined as follows:
\begin{align}
  (\text{grad}\, s)(i, j) &=  s_j-s_i \quad \text{for $\{i,j\} \in E$} \label{grad}, \\
  (\text{curl}\, A)(i, j, k) &= A_{ij} + A_{jk}+A_{ki}\quad \text{for $\{i,j,k\} \in T(E)$} \label{curl},
\end{align}
where $s$ is a potential function, $T(E)$ is the set of triangles in the graph, and $\{\{i, j, k\}: \{i, j\}, \{j, k\}, \{k, i\} \in E\}$.
Using these operators, the space of edge flow $\mathcal{A}$ is 
orthogonally decomposed into three subspaces,
\begin{align}
  \mathcal{A}  
  &= \underbrace{\text{im}(\text{grad})}_{\text{gradient flow}} \oplus \underbrace{\text{ker}(\Delta_1)}_{\text{harmonic flow}} \oplus \underbrace{\text{im}(\text{curl}^*)}_{\text{curl flow}},
\end{align}
where ker($\Delta_1$) = ker(curl) $\cap$ ker(div) and $\text{curl}^*$ is the adjoint operator of the curl.
With a Euclidean inner product in  the space $\mathcal{A}$, $\langle X,W\rangle = \sum_{ \{i,j\} \in E} X_{ij}Y_{ij}$,
we define an optimisation problem:
\begin{align}
  \min_s  \lVert   \text{grad}\ s - A \rVert_2 = \min_s \left[ \sum_{\{i,j\} \in E} \left( (s_j - s_i) - A_{ij} \right)^
  2 \right]  \label{eq:optimization_problem_general}.
\end{align}
This is equivalent to an $l_2$-projection of $A$ onto im(grad).
Then, the solution of the optimisation problem satisfies the following normal equation \cite{Jiang2011}:
\begin{align}
  \Delta_0 s = -  \text{div} A,
\end{align}
where $\Delta_0$ is the graph Laplacian of graph $G$ and the divergence is (div $A$)($i$) = $\sum_{j \text{ s.t. } \{i,j\} \in E} A_{ij}$. 
Potential $s$ with the minimal norm is given by,
\begin{align}
  s = - \Delta_0^{\dagger} \text{div} A,  \label{eq:scalar_potential}
\end{align}
where $\dagger$ denotes the Moore-Penrose inverse.
Similarly, vector potential $\Phi$ of curl flow is derived as
\begin{align}
  \Phi =(\mathrm{curl} \circ \mathrm{curl}^{*} )^{\dagger} \mathrm{curl} A. \label{eq:vector_potential}
\end{align}

The OD matrix determines the edge flow for every pair of nodes, corresponding to a complete graph.
It should be noted that no edge between ${i,j}$ in graph $G$ means that the flow $A_{ij}$ between them is undefined or unavailable, and does not mean zero movement, $A_{ij} = 0$.
In this case of a complete graph,
$\text{dim}(\text{ker} (\Delta_1) )$ = 0 holds, and matrix $A$ is decomposed into only two parts: gradient and curl flows.
Furthermore, the scalar potential in \eqref{eq:scalar_potential} and the vector potential in \eqref{eq:vector_potential} are simplified as follows:
\begin{align}
  s_i &= -\frac{1}{N}\sum_{j=1}^N A_{ij},\\
  \Phi_{ijk} & = \frac{1}{N}(A_{ij} + A_{jk} + A_{ki}),
\end{align}
where $N$ is the number of nodes. Using these potentials,
the net flow $A$ is uniquely decomposed as
\begin{align}
  A_{ij} & = (\text{grad}\, s)(i, j)  +  (\text{curl}^*\, \Phi)(i, j) \notag \\
         &= (s_j - s_i) + \sum_k \Phi_{ijk}.
\end{align}

\subsection*{Datasets}
\subsubsection*{London} 
We used a person trip dataset from home to workplace in 2011,
obtained from UK Data Service \cite{TripDataLondon}.
The OD matrix denotes the number of commuters aggregated by the middle layer super output area (MSOA) in the census 2011.
The shapefile of the MSOAs is obtained from Office for National Statistics \cite{ShapeDataLondon}.
The dataset covers the MSOAs in England and Wales.
In this paper, we selected the trips only among MSOAs in Greater London.
The resultant matrix contains 2.9 million trips among 983 MSOAs.

The trips in the dataset are categorised by main transport methods
and we selected the following two types of transport methods:
``Public transport''  includes the trips by underground, metro, light rail, tram, train, Bus, minibus, and coach.
``Private car'' includes the trips by driving a car, taxi, motorcycle, scooter, or moped, including their passengers.

\subsubsection*{Tokyo} 
We used  datasets from successive person-trip surveys from 1988 to 2018 in the Tokyo metropolitan area \cite{TripDataTokyo}.
The datasets were categorised according to the aims of trips, and home-work trips were selected.
The OD matrix denotes the number of commuters aggregated by the middle-sized geographical zones.
The middle-sized zone is essentially equivalent to a municipal district, except that some zones contain several districts in rural areas.
The zones have been altered by municipal mergers and dissolutions from 1988 to 2018,
and the target regions of the surveys have been extended.
We selected the areas covered by all surveys from 1988 to 2018
and have a surjective mapping from the zones in 2018, to make shapefiles before 2018 (shapefile is available only for the last survey in 2018).
Several peripheral areas in the Ibaragi, Chiba, Kanagawa, and  Saitama provinces were excluded.

The resultant matrix contains 
11.77 million trips among 121 zones in 2018, 
11.74 million trips among 120 zones in 2008, 
10.97 million trips among 114 zones in 1998, 
9.97  million trips among 106 zones in 1988.

\subsubsection*{Core-based statistical area (CBSA)  in the United States} 
We used LEHD Origin-Destination Employment Statistics (LODES) datasets in 2018 \cite{TripDataUS}. 
The datasets contain the number of jobs for each pair of residential places and workplaces at the census block level. We aggregated the data into the census tract level and analysed the OD matrix of commute trips for each core-based statistical area (CBSA) defined by the U.S. Office of Management and Budget \cite{CBSADefinition}. The shapefiles of the census tracts were obtained from 2019 TIGER/Line shapefiles \cite{ShapeDataUS}.

There are 930 CBSAs in 2018, when those in Alaska and Puerto Reco are excluded \cite{CBSAList}.
CBSAs are classified into metropolitan statistical areas (MSAs) and micropolitan statistical areas ($\mu$SAs), depending on whether the population is larger than 50,000.

The population of a CBSA is computed by adding  those of counties that belong to the CBSA. County populations were taken from \cite{CBSAPopulation}.

\clearpage
\setcounter{figure}{0} 
\setcounter{table}{0} 
\renewcommand{\thefigure}{S\arabic{figure}}
\renewcommand{\thetable}{S\arabic{table}}

\section*{Supplementary Text}
\section*{1. Benchmark test using synthetic OD matrix}

\begin{figure}[tb]
  \includegraphics[width=\linewidth]{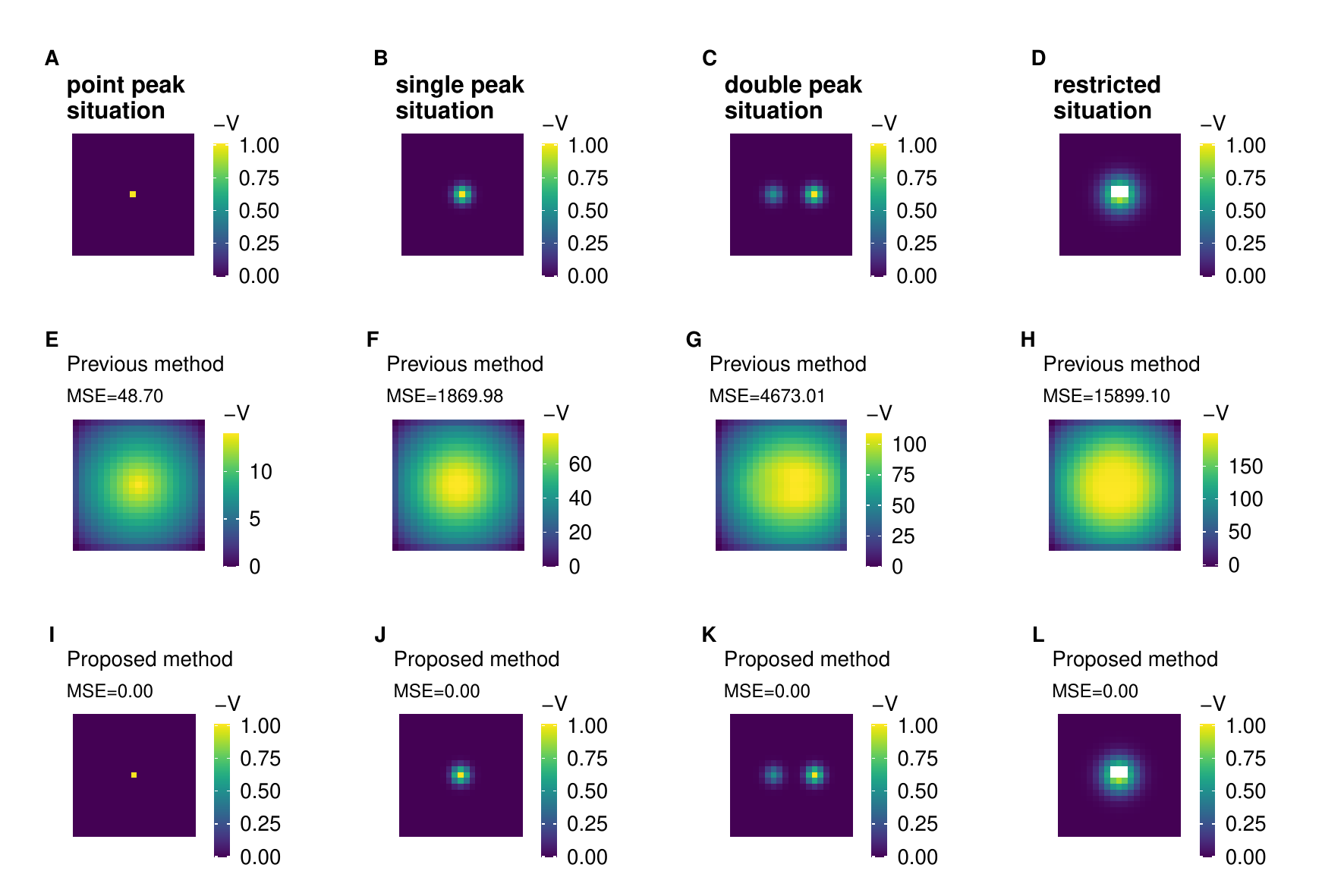}
  \caption{
    \textbf{Potential identification for conceptual situations.}
    (\textbf{A}) 
  \textit{point peak} situation represents an artificial monocentric city.
There is the attractive location at the centre, to which people move from all other locations.
    (\textbf{B}) 
\textit{single peak} situation is similar to the \textit{point peak} situation, but the central place has some spatial extents.
    (\textbf{C}) 
\textit{double peak} situation represents a polycentric city, in which there are two attractive places.
    (\textbf{D}) 
    In \textit{restricted} situation, the potential in some locations is undefined (white cell), because it is, for example, a restricted area or a lake.
These locations can not be the origin or destination of flow.
    (\textbf{E,F,G,H}) 
    The potential identified by the method proposed in \cite{Mazzoli2019b} in each situation shown in the upper panel.
    (\textbf{I,J,K,L}) 
    The potential identified by the method by Hodge-Kodaira decomposition in each situation.
  }
  \label{fig:synthetic}
\end{figure}

To confirm the validity of potential extraction methods, we performed benchmark tests.
For the tests, the OD matrix $M$ was synthetically derived from a given potential $V$:
\begin{equation}
  M_{ij} = [ - (V_j - V_i) ]_{+}, \label{eq:synthetic_gradient}
\end{equation}
where $[x]_+ = \max(0,x)$ is rectifier to ensure positive trips.
For the synthetic OD matrix, we treat $V$ as the ``ground truth'' of the potential. %
By comparing with this true potential $V$,
we validated the extracted potential $\hat{V}$ from the OD matrix $M$
and calculated the mean squared error (MSE):
\begin{equation}
  \text{mean squared error} = \frac{1}{N} \sum_{i = 1}^N   (\hat{V}_i - V_i)^2. \label{eq:MSE}
\end{equation}
In the comparison, potential is standardised so that its maximum value matches with the reference value ($V=0$).

Figure \ref{fig:synthetic} shows the benchmark results;
the top panels show typical urban structures represented by the potential $V$
and the middle and bottom panels show those obtained from the previous method in \cite{Mazzoli2019b} and the proposed method, respectively.
This visualises whether each of the two methods correctly recovers the ``true'' structures from the synthetic OD matrix.

The first \textit{point peak} situation represents an ideal monocentric city (Fig. \ref{fig:synthetic}A).
There is an attractive location at the centre and the potential of the other locations is equal to the reference value ($V$ = 0).
In this situation, people gather at the single point at the centre and the flow is seen as a star network.
Under this condition, the potential $\hat{V}$ estimated by the previous method in \cite{Mazzoli2019b} is peaked at the same centre but has a broader distribution.
This suggests that the area near the centre is differentiated by the extraction potential $\hat{V}$ from more peripheral areas.
This is inconsistent with the ground truth of the city structure, in which all locations are identical except for the central point.

Next, in the \textit{single peak} situation (Fig. \ref{fig:synthetic}B), the attractive place has some spatial extents.
In this condition, the potential $\hat{V}$ extracted by the method in \cite{Mazzoli2019b}  still has a wider distribution; therefore, the central area by $\hat{V}$ appears larger than its actual size.

The \textit{double peak} situation represents a polycentric city (Fig. \ref{fig:synthetic}C).
There are two attractive places that draw the flow of people from the other locations.
The place on the right-hand side is more attractive than the one on the left-hand side, as shown by their potential values; thus, the right-hand side is the main centre and the left-hand side is a sub-centre.
The extracted potential $\hat{V}$ in \cite{Mazzoli2019b} has only a single-peak broad distribution;
thus, it is difficult observing the clear polycentric structure.
This misidentification is caused by the conversion process from the place-to-place flow to 2d vector field, which is intrinsic to the method.
In the Tokyo metropolitan area, for example, \textit{Kawasaki} city is known as a sub-centre \cite{Li2018} to which people commute. At the same time, many residents in the area commute to the largest central area around \textit{Chiyoda} city.
This is a common situation in metropolitan area, 
which is often defined as an urban centre and its commuter hinterland,
such as core-based statistical areas in the US or travel-to-work areas in UK.
In this case, the averaged vector at the \textit{Kawasaki} city will be directed toward the largest centre, and then the sub-centre becomes hidden.
In fact, the potential shown in \cite{Mazzoli2019b} for the Tokyo metropolitan area has no peak at the \textit{Kawasaki} city and other known sub-centres;
therefore, the previous method would be unsuitable for discussing polycentric structures within metropolitan areas.

The \textit{restricted} situation in Fig. \ref{fig:synthetic}D is similar to the \textit{single peak} situation,  with some locations near the centre are labelled as \textit{NA}.
Here, the \textit{NA} label indicates the potential is undefined because the location is a non-land cell, such as rivers, lakes, and seas, or a restricted area by law.
For example, some historical cities have developed around palaces or castles, which often have restricted areas.
The \textit{NA} locations cannot be the origin or destination of a flow.
The potential obtained by the previous method spreads over the \textit{NA} locations, which has also been observed in real cities \cite{Mazzoli2019b}.
Furthermore, it identifies those locations as a part of a central area.
The potential at a non-land cell or restricted area would be difficult to interpret.

These observed deviations from the true potential $V$ are quantitatively measured in terms of mean squared error.
Although the large errors obtained via the previous method are partly caused by inconsistencies in the generating process of the OD matrix by equation \eqref{eq:synthetic_gradient},
the concepts of the examined situations are generic, independent of the specific equation.

In contrast to the previous method, Hodge-Kodaira decomposition perfectly recovers the true potentials without any error in all situations.
It does not assign any potentials at non-land cells or restricted areas, that is those labelled as  \textit{NA}.
Therefore, we can identify the urban structures, such as the location of centre in city, its area size, or the number of centres, as represented by the potential.

It is noted that these benchmark examples are not too disadvantageous for the previous method. It is actually advantageous: the method requires that flows are given at grid points as presented in this benchmark. However, human flow datasets are usually aggregated by administrative units in survey-based collections or Voronoi polygons of cell towers in call detail records (CDRs) of mobile phones, and some resampling treatments to grid points are needed.
In contrast, the Hodge-Kodaira decomposition is applicable to the OD matrix aggregated by any shape of geographical zones.

\section*{2. Ranking of potential in the Tokyo metropolitan area from 1988 to 2018}
Table \ref{tbl:top20_tokyo} shows the top 20 zones by  negative potential $-V$,
using the commuter datasets of successive person-trip surveys from 1988 to 2018 in the Tokyo metropolitan area.

Over 30 years, \textit{Chiyoda} city --- the Imperial Palace and its surrounding areas ---  has been at the top of the potential.
Its neighbouring cities, such as \textit{Minato}, \textit{Chuo}, \textit{Shinjuku}, and \textit{Shibuya}, had occupied the top five ranks by the potential over the years.

Some zones, such as \textit{Naka} ward in \textit{Yokohama} city, \textit{Kawasaki} ward in \textit{Kawasaki} city, and \textit{Sumida} city, have declined in the ranking from 1988 to 2018.
On the other hand, some zones, such as  \textit{Tachikawa} and \textit{Akishima} cities and \textit{Omiya} ward in \textit{Saitama} city, have raised their rankings.

\begin{sidewaystable}[ht]
\centering
\scalebox{0.85}{
\begin{tabular}{p{10em}rp{10em}rp{10em}rp{10em}r}
  \toprule
  \multicolumn{2}{l}{2018} & \multicolumn{2}{l}{2008} &\multicolumn{2}{l}{1998} & \multicolumn{2}{l}{1988} \\ 
 Zone & $-V$ & Zone & $-V$ & Zone & $-V$ &Zone & $-V$ \\ 
 \midrule
Chioda City & 6554 & Chioda City & 7170 & Chioda City & 5916 & Chioda City & 6808 \\ 
  Minato City & 5367 & Minato City & 6580 & Minato City & 4952 & Chuo City & 5251 \\ 
  Chuo City & 3337 & Chuo City & 4454 & Chuo City & 4412 & Minato City & 5164 \\ 
  Shinjuku City & 2927 & Shinjuku City & 3088 & Shinjuku City & 2999 & Shinjuku City & 2854 \\ 
  Shibuya City & 2237 & Shibuya City & 2280 & Shibuya City & 2101 & Shibuya City & 1696 \\ 
  Shinagawa City & 1358 & Shinagawa City & 1317 & Taito City & 1142 & Taito City & 1236 \\ 
  Koto City & 1221 & Koto City & 992 & Shinagawa City & 1054 & Naka Ward, Yokohama City & 849 \\ 
  Bunkyo City & 695 & Taito City & 878 & Toshima City & 875 & Shinagawa City & 665 \\ 
  Taito City & 616 & Bunkyo City & 824 & Naka Ward, Yokohama City & 793 & Toshima City & 647 \\ 
  Nishi Ward, Yokohama City & 578 & Naka Ward, Yokohama City & 704 & Bunkyo City & 783 & Kawasaki Ward, Kawasaki City & 633 \\ 
  Naka Ward, Yokohama City & 513 & Toshima City & 682 & Koto City & 658 & Bunkyo City & 625 \\ 
  Atsugi City with Aikawa and Kiyokawa Areas & 327 & Nishi Ward, Yokohama City & 561 & Nishi Ward, Yokohama City & 610 & Nishi Ward, Yokohama City & 402 \\ 
  Chuo Ward, Chiba City & 286 & Chuo Ward, Chiba City & 372 & Kawasaki Ward, Kawasaki City & 605 & Sumida City & 322 \\ 
  Toshima City & 272 & Kawasaki Ward, Kawasaki City & 330 & Chuo Ward, Chiba City & 585 & Atsugi City with Aikawa and Kiyokawa Areas & 191 \\ 
  Mihama Ward, Chiba City & 217 & Atsugi City with Aikawa and Kiyokawa Areas & 283 & Sumida City & 343 & Chiba City & 183 \\ 
  Kawasaki Ward, Kawasaki City & 214 & Sumida City & 233 & Atsugi City with Aikawa and Kiyokawa Areas & 207 & Koto City & 129 \\ 
  Omiya Ward, Saitama City & 180 & Mihama Ward, Chiba City & 137 & Yokosuka City & 122 & Meguro City & 58 \\ 
  Tachikawa and Akishima Cities & 117 & Omiya Ward, Saitama City & 116 & Ota City & 106 & Ota City & 47 \\ 
  Sumida City & 85 & Tachikawa and Akishima Cities & 42 & Meguro City & 89 & Narita City with Tomisato and Sakae Towns & 37 \\ 
  Saiwai Ward, Kawasaki City & 80 & Yokosuka City & 39 & Narita City with Tomisato and Sakae Towns & 85 & Yokosuka City & 35 \\ 
   \bottomrule
\end{tabular}
}
\caption{Top 20 zones of potentials in Tokyo metropolitan area from 1998 to 2018.} 
\label{tbl:top20_tokyo}
\end{sidewaystable}

\section*{3. Percentage of gradient components in London and Tokyo.}
The percentage $R^2$ was evaluated  by the transport methods in the London case (Table \ref{tbl:LondonPercentage}) and  by year in the Tokyo case (Table \ref{tbl:TokyoPercentage}).

\begin{table}[ht]
\centering
\begin{tabular}{lr}
  \toprule
Transport Method & Percentage \\ 
  \midrule
All & 52.7 \\ 
  Public transport & 63.7 \\ 
  Private car & 21.5 \\ 
   \bottomrule
\end{tabular}
\caption{Percentage $R^2$ of gradient component of home-work trips in London.} 
\label{tbl:LondonPercentage}
\end{table}

\begin{table}[ht]
\centering
\begin{tabular}{rr}
  \toprule
Year & Percentage \\ 
  \midrule
1988 & 42.4 \\ 
  1998 & 41.3 \\ 
  2008 & 40.8 \\ 
  2018 & 37.4 \\ 
   \bottomrule
\end{tabular}
\caption{Percentage $R^2$ of gradient component of home-work trips in Tokyo metropolitan area.} 
\label{tbl:TokyoPercentage}
\end{table}

\section*{4. Distance deterrence effect.}
Distance is obviously an important factor of human mobility.
This factor has been considered as a \textit{distance-deterrence function} $f(d)$ in spatial interaction models, such as the gravity model.
We here integrate the function into the Hodge-Kodaira decomposition.

We rewrite the optimization problem in equation (\ref{eq:optimization_problem}) as
\begin{align}
  & \min_s \left[ \sum_{\{i,j\} } \left[ f(d_{ij})(s_j - s_i) - A_{ij} \right]^2 \right] \nonumber\\
  &  = 
  \min_s \left[ \sum_{\{i,j\} } |f(d_{ij})|^2 \left[ (s_j - s_i) - \frac{A_{ij} }{f (d_{ij})} \right]^2 \right] \nonumber \\
  &= 
  \min_s \left[ \sum_{\{i,j\} } w_{ij} \left[ (s_j - s_i) - A^w_{ij} \right]^2 \right],
\end{align}
where
\begin{align*}
  A^{w}_{ij} &:=  A_{ij} / f(d_{ij})\\
  w_{ij} &:= |f(d_{ij})|^2.
\end{align*}
With a weighted Euclidean inner product in the space $\mathcal{A}$,
$\langle X,W\rangle = \sum_{ \{i,j\} } w_{ij} X_{ij}Y_{ij}$,
the problem is equivalent to an $l_2$-projection of $A$ onto im(grad)
and the minimal norm solution $s^w$  is given by,
\begin{align}
  s^w_i = - \frac{1}{N} \text{div} (A)(i) 
  = - \frac{1}{N} \sum_j^N w_{ij} A^w_{ij}
  = - \frac{1}{N} \sum_j^N f(d_{ij}) A_{ij}. \label{eq:distance-integrated}
\end{align}

\begin{figure}[tb]
  \includegraphics[width=\linewidth]{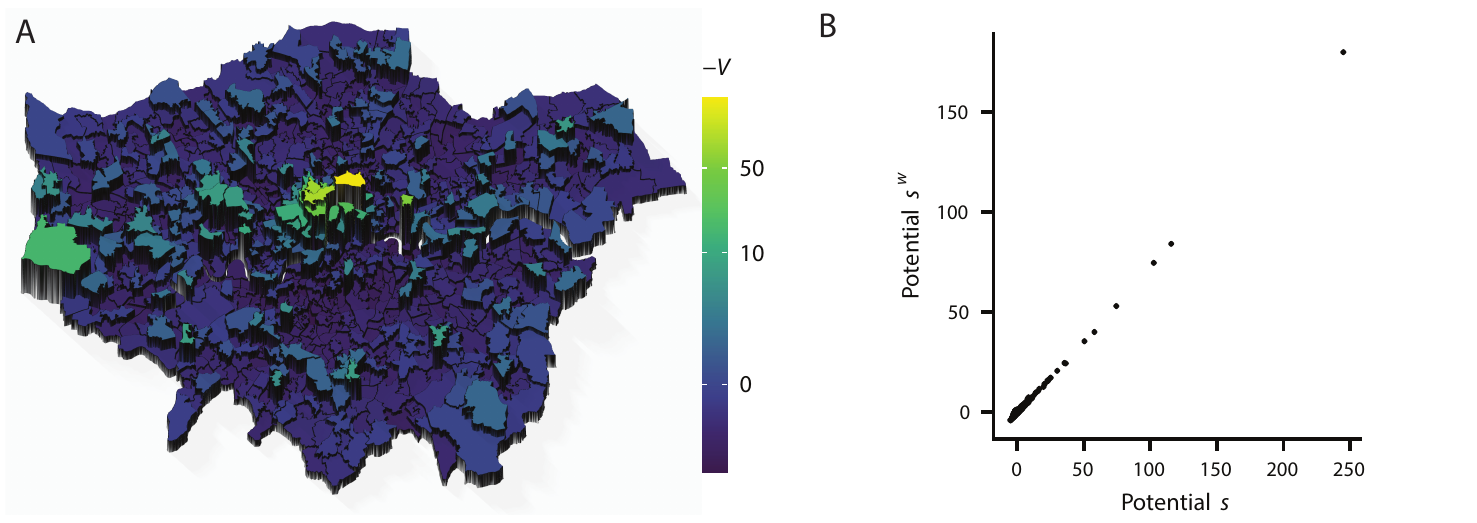}
  \caption{
  \textbf{Distance-integrated potential $s^w$ in the London case.}
    (\textbf{A}) 
    The potential is calculated by equation (\ref{eq:distance-integrated}), using the same 
    home-work trip data in Figure \ref{fig:PotentialLondon}.
    (\textbf{B}) 
    Scattering plot between the potential without the distance-deterrence effect, $s$ and with the effect, $s^w$.
  }
  \label{fig:distance-integrated-London}
\end{figure}

Figure \ref{fig:distance-integrated-London} shows this distance-integrated potential $s^w$ in the London case, with a specified  form of the distance-deterrence function, $ f(d) = \exp(-d_{ij}/D)$.
The parameter $D$ was 31.12 km, determined by a linear regression with the formula $A_{ij} = \text{constant} \cdot \exp(-d_{ij}/D)$, using the Euclidean distance $d_{ij}$ between the centroids of the zones.
As shown in Figure \ref{fig:distance-integrated-London}(A),
the landscape of the distance-integrated potential $s^w$ is similar to that of potential $s$ without distance-integration.
Figure \ref{fig:distance-integrated-London}(B) shows the scattering plot between $s$ and  $s^w$.
The points lie on a line, indicating a sublinear scaling of $s^w$ to $s$.
This means that in the London case, the distance-deterrence effect rescales the potential $s$ uniformly, and does not alter the relative rankings among the places by the potential.

\end{document}